\documentclass[final,3p,times,twocolumn]{elsarticle}
\usepackage{amssymb,amsmath}
\begin{document}

\begin{frontmatter}



%\title{Enhancing the effective Fe abundance in UHECR sources}

\title{Enhancing the Relative Fe-to-Proton Abundance \\ in Ultra-High-Energy Cosmic Rays}

%\title{The Effect of a Maximum Source Energy Distribution on the Fe-to-proton ratio in Ultra-High-Energy Cosmic Rays}


\author{Carl Blaksley}
\author{Etienne Parizot}

\address{Laboratoire Astroparticule et Cosmologie (APC), Universit\'e Paris 7/CNRS, 10 rue A. Domon et L. Duquet, 75205 Paris Cedex 13, France}

\begin{abstract}

We study a generic class of models for ultra-high energy cosmic ray (UHECR) phenomenology, in which the sources accelerate protons and nuclei with a power-law spectrum having the same index, but with different values for the maximum proton energies, distributed according to a power-law. We show that, for energies sufficiently lower than the maximum proton energy, such models are equivalent to single-type source models, with a larger effective power law index and a heavier composition at the source. We calculate the resulting enhancement of the abundance of nuclei, and find typical values of a factor 2--10 for Fe nuclei. At the highest energies, the heavy nuclei enhancement ratios become larger, and the granularity of the sources must also be taken into account. We conclude that the effect of a distribution of maximum energies among sources must be considered in order to understand both the energy spectrum and the composition of UHECRs, as measured on Earth.

\end{abstract}


\end{frontmatter}

% \linenumbers

\section{Introduction}
\label{sec:intro}

The average nuclear composition of ultra high energy cosmic rays (UHECR) at their sources is one of the key ingredients in their phenomenology. It has been shown that a mixed composition, that is to say the presence of nuclei in addition to protons, implies a harder source spectrum, typically in $E^{-x}$ with $x \simeq 2.2-2.3$ as opposed to $x\simeq 2.6-2.7$ in the case of a pure proton composition (depending on the cosmological evolution of the source power)\cite{Allard+08}. In addition, a mixed composition implies that the energy at which the extra-galactic component of the total UHECR flux becomes larger than
 the Galactic component should be somewhere around the so-called ankle, i.e. $\sim 5\cdot10^{18}$~eV, while this transition would be at a lower energy in the pure proton case \cite{BerGri88,Allard+05,Allard+07,Aloisio+07}.

Any accurate description of the UHECR data should reproduce both the measured spectrum \emph{and} composition in a consistent way, and, while the measurement of the UHECR composition at Earth remains a difficult observational task, important progress has been made in the recent years.
Notable results include those of the Pierre Auger Observatory \cite{AugerCompo09,AugerXmax10}, which has provided hints that the composition becomes heavier and heavier above $\sim 10^{19}$~eV (assuming the general validity of hadronic interaction models). In contrast to this, other experiments such as HiRes \cite{Abbasi+04} and Telescope Array \cite{Kawai+08,Thomson+10} have shown results which are compatible with pure proton scenarios.

Propagation effects are known to modify the composition of UHECRs, as the energetic nuclei are photo-dissociated in interactions with background photons.
 Horizon analyses show that Fe nuclei and protons with energies above $\sim 6\cdot10^{19}$~eV can propagate over roughly the same distance without losing a significant
 fraction of their total energy, while intermediate mass nuclei are suppressed at shorter distances \cite{Harari+06,Allard+08}. As a result UHECRs, at these high energies, should be dominated by either protons, Fe (or sub-Fe) nuclei, or a combination of the two. Therefore, the Auger results on UHECR 
composition can be understood if the proton component is cut at the source at a relatively low energy, around $10^{19}$~eV, and Fe nuclei are accelerated 
up to higher energies, eventually dominating the overall spectrum. This would be natural, for instance, in a scenario where the different nuclei reach a 
maximum energy at the source that is proportional to their charge, $Z$.

Although it has been shown that it is indeed possible to fit the UHECR energy spectrum in such a scenario \cite{Allard+08}, this requires a source composition which is richer in Fe nuclei than would be expected from a simple extrapolation of the low energy cosmic ray source composition. This can be achieved, in principle, by invoking a source environment which is richer in Fe nuclei, or an acceleration mechanism which somehow discriminates nuclei on the basis of mass. In this paper, we propose another mechanism to produce a heavier \emph{effective} composition at the source, independent 
of the acceleration model. We assume only that the maximum energy reached by the particles accelerated in a given source is not universal and that the distribution of
 sources with respect to their maximum energy follows a power-law. The contribution of all sources is then equivalent to a scenario where identical sources 
not only inject UHECRs with a softer spectrum, as already shown by \cite{KacSem06}, but also with a heavier source composition. We then compute 
the Fe-to-proton enhancement ratio, $\eta_{\mathrm{Fe}}$, as a function of the source parameters.

% In section \ref{sec:source} we will.... Section \ref{sec:ratio} will..., and section \ref{sec:conclu} we will...

%\section{Source $E_{max}$ Distributions and the Proton-Fe Ratio}
\section{Source $E_{max}$ Distributions and Resulting Spectra}
\label{sec:source}

%\subsection{UHECR spectrum including sources with a distribution of $E_{\max}$.}

For most proposed cosmic ray sources the maximum energy which can be reached is limited by the ability of the source to contain particles in the acceleration region, i.e. by particle rigidity, defined as $R \equiv p/q$. In
the relativistic limit applicable to UHECRs, this is proportional to $E/Z$, where $Z$ is the charge of the accelerated nuclei. In such a case, unless other mechanisms come into play to limit the energy of specific nuclei (such as photo-dissociation processes), the maximum energy, $E_{\max}^{(i)}$, of nuclei of type $i$ at the source is simply proportional to their charge, $Z_{i}$:
\begin{equation}
E_{\max}^{(i)} = Z_{i}\times E_{\max}^{(\mathrm{p})}\,,
\label{eq:emaxrelation}
\end{equation}
where $E_{\max}^{(\mathrm{p})}$ is the proton maximum energy.

For simplicity, UHECR models usually assume that all cosmic ray sources are identical, having the same spectrum extending up to the same energy. However, it is clear that the maximum energy will differ among sources depending on their individual properties, notably their size, magnetic field strength, intrinsic power, age, etc. In order to avoid introducing extra free parameters the simplest assumption is to assume a power-law distribution for the number of sources as a function of $E_{\max}^{(p)}$ (as in \cite{KacSem06}):

\begin{equation}
\label{eq:numbersources}
n_{\mathrm{sources}}(E_{\max}^{(p)})= n_{0}\left(\dfrac{E_{\max}^{(p)}}{E_{0}}\right)^{-\beta} H\left(E_{\sup}-E_{\max}^{(p)}\right),
\end{equation}
where we have introduced $E_{\sup}$ as the highest possible proton energy in any source and $H(x)$ is the Heaviside step function. The parameter $E_{0}$ is an arbitrary reference energy, which has been introduced here simply to clarify the dimensionality of the various quantities.
$\beta$ is expected to be a positive number, as the number of sources able to accelerate nuclei to high energies presumably decreases with increasing energy.

%It is important to note that $E_{sup}$ serves not only to avoid a divergence in the total number of sources, but also physically represents the limit of particle acceleration mechanisms if such a limit exists. 

We assume that the individual sources each produce a power-law spectrum of UHECRs with the same spectral index, $x$. The number of nuclei of type $i$ injected per second and per unit energy by a given source is then:

\begin{equation}
 \label{eq:injection}
Q_{i}(E) \equiv \dfrac{\mathrm{d}^{2}N}{\mathrm{d}E\,\mathrm{d}t}=Q_{0}\,\alpha_{i}\left(\dfrac{E}{E_{0}}\right)^{-x} H(E_{\max}^{(i)}-E),
\end{equation}
where $\alpha_{i}$ is the abundance of nuclear species $i$, with $\sum \alpha_{i} = 1$.

Assuming that the sources are homogeneously distributed over space, the total number of cosmic rays injected per unit time, per unit energy, and per unit volume is then given by:

\begin{equation}
q_{i}(E) = \int_{0}^{\infty} Q_{i}(E)\times n_{\mathrm{sources}}(E_{\max}^{(p)})\,\mathrm{d}E_{\max}^{(p)}.
\end{equation}

Replacing from Eqs.(\ref{eq:numbersources}) and~(\ref{eq:injection}), one gets:
\begin{equation}
q_{i}(E) = Q_{0}n_{0}\alpha_{i}\left(\frac{E}{E_{0}}\right)^{-x}\,\int_{0}^{E_{\sup}} \left(\frac{E_{\max}^{(p)}}{E_{0}}\right)^{-\beta}H(Z_{i}E_{\max}^{(p)} - E)\,\mathrm{d}E_{\max}^{(p)},
\end{equation}
 which integrates (for $\beta \neq 1$) into:
%\begin{equation}
%q_{i}(E) = \frac{Q_{0}n_{0}E_{0}}{\beta-1}\,\alpha_{i}Z_{i}^{\beta-1}\left(\frac{E}{E_{0}}\right)^{-x}\left[\left(\frac{E}{E_{0}}\right)^{1-\beta} - \left(\frac{Z_{i}\,E_{\sup}}{E_{0}}\right)^{1-\beta}\right].
%\label{eq:effectiveInjection}
%\end{equation}
\begin{equation}
q_{i}(E) = \frac{Q_{0}n_{0}E_{0}}{\beta-1}\,\alpha_{i}Z_{i}^{\beta-1}\left(\frac{E}{E_{0}}\right)^{-x-\beta+1}\left[1 - \left(\frac{E}{Z_{i}\,E_{\sup}}\right)^{\beta-1}\right].
\label{eq:effectiveInjection}
\end{equation}

NB: if $\beta = 1$, there are the same number of sources in each decade of $E_{\max}$, leading to a logarithmic divergence in the integral if $E_{\sup}$ tends to infinity. In most natural cases, however, one would expect $\beta$ to be larger than 1.

\section{Effective Spectrum and Composition in the limit of $E\ll E_{\sup}$}

For values of $E_{\sup}$ much larger than $E$ (and $\beta > 1$), Eq.~(\ref{eq:effectiveInjection}) is equivalent to a single source power-law distribution given by:
\begin{equation}
q_{i}(E) = q_{0}\,\alpha_{i}Z_{i}^{\beta-1}\left(\frac{E}{E_{0}}\right)^{-x-\beta+1}.
\label{eq:effectiveSpectrum}
\end{equation}

In other words, the simple model described here is equivalent to the usual ``universal source model'' with the effective source spectral index
\begin{equation}
x_{\mathrm{eff}} = x + \beta - 1,
\label{eq:xEff}
\end{equation}
and with a modified source composition corresponding to the effective nuclear abundances:
\begin{equation}
\alpha_{i,\mathrm{eff}} = \alpha_{i}\times Z_{i}^{\beta-1}.
\label{eq:alphaEff}
\end{equation}

The effect of a distribution of $E_{\max}$ values among sources is thus a modification of both the energy spectral index \emph{and} the abundances of nuclei in a correlated way through the parameter $\beta$. This shows that these two aspects of UHECR phenomenology are not independent.

\begin{figure}[ht!]
\begin{center}
 \includegraphics[width=1.1\columnwidth]{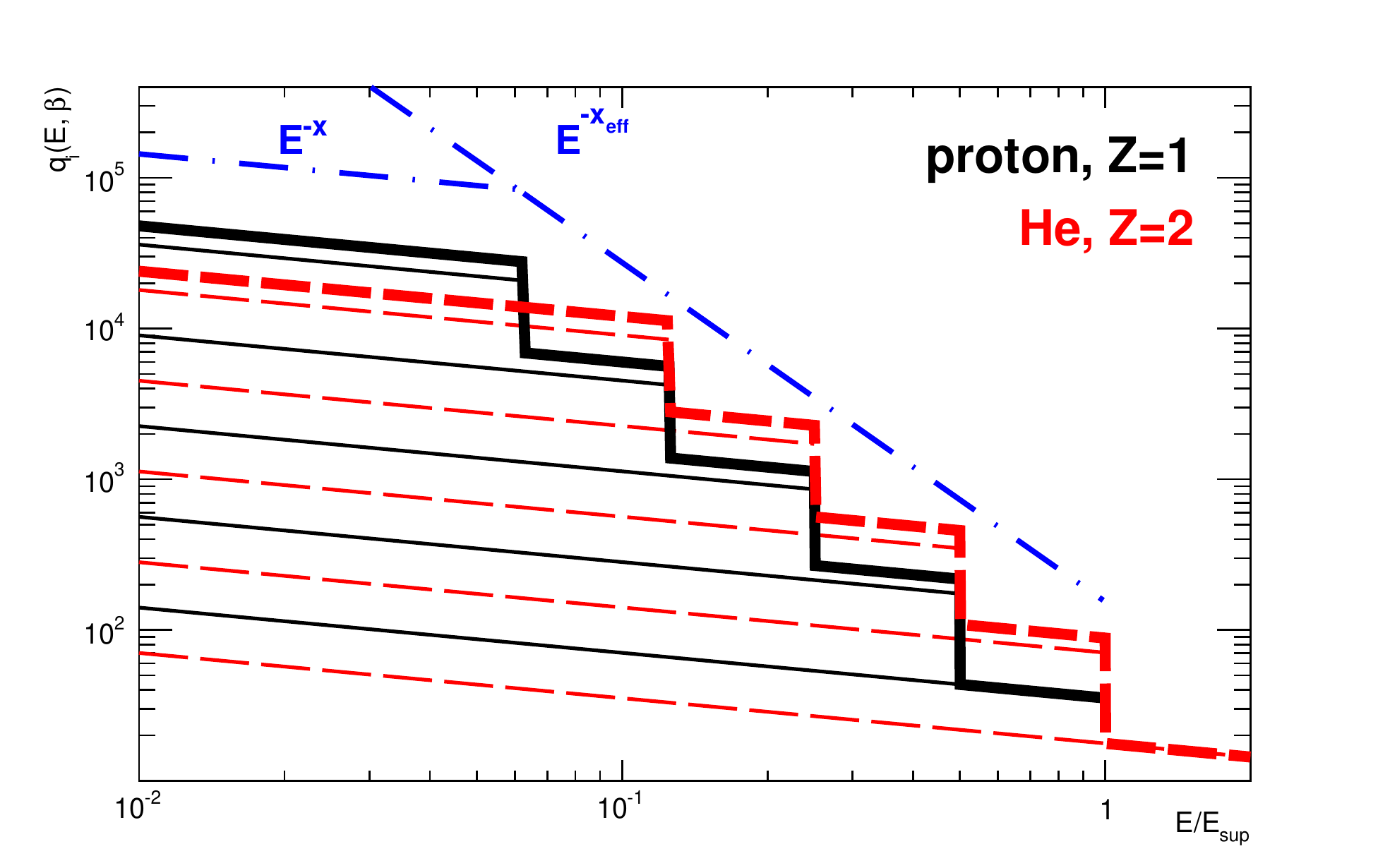}
\caption{An illustration of the effect of an $E_{\max}$ distribution on the ratio of different nuclei in UHECRs. Here, we consider an artificial distribution of sources with 5 different maximum energies, as explained in the text. The solid lines are the proton spectra for each source, while the dashed lines denote the He spectra, extending to twice higher energies. The total effective spectrum for each species (before propagation) is given by the bold lines. In this example, even though the abundance of protons at the source is higher, in the UHECR region the spectra of the heavier nuclei are enhanced due to the distribution of the number of sources with respect to $E_{\max}$.}
\label{fig:distributions}
\end{center}
\end{figure}

This behavior is easily understood by considering the simple example, say, of 5 discrete sources, as illustrated in Fig.~\ref{fig:distributions}. In this plot, the plain lines show an identical proton injection spectrum, extending up to a maximum energy that is different for each source. The dashed lines show the corresponding helium injection spectra, with an assumed abundance ratio of $\alpha_{\mathrm{He}}/\alpha_{\mathrm{p}} = 0.5$, and a maximum energy twice as large for each source, as per Eq.~(\ref{eq:emaxrelation}). The bold lines are then simply the sum of all 5 contributions, showing an enhancement of the relative helium abundance. In this illustrative example, while the He nuclei are less abundant than the protons in each individual source, they dominate the overall flux injected at high energy.

Fig.~\ref{fig:distributions} also shows how the original source spectrum steepens as fewer and fewer sources contribute to the flux at higher and higher energies. Below the $E_{\max}$ of the least energetic source, the sum spectrum obviously has the same form as the individual source spectra. Above that energy, and up to $\sim E_{\sup}$ (i.e. the $E_{\max}$ of the most energetic source), the resulting spectrum is a new power-law with the larger spectral index, $x_{\mathrm{eff}}$, as shown by the dashed line fit.

In an actual astrophysical context, extensive propagation studies have determined the effective spectral index needed to reproduce the observed UHECR energy spectrum for a given assumed source composition (see Sect.~\ref{sec:intro}). In the case of pure proton sources the composition effect demonstrated above is obviously irrelevant, but for a mixed composition scenario Eq.~(\ref{eq:alphaEff}) shows that the effective composition to be used in the models is systematically richer in heavy nuclei than the individual source composition, as soon as $\beta > 1$. Given Eq.~(\ref{eq:xEff}), this condition is equivalent to saying that the effective source spectrum is steeper than the intrinsic one ($x_{\mathrm{eff}} > x$). For instance, if the source spectral index is $x = 2.0$, as expected from standard diffusive shock acceleration, the effective source spectrum needed to reproduce the data in the case of a mixed composition model, namely $x_{\mathrm{eff}} \simeq 2.3$, requires a source distribution index $\beta \simeq 1.3$. This in turn implies that the effective Fe nuclei abundance is larger than the Fe abundance in individual sources by a factor $Z_{\mathrm{Fe}}^{\beta - 1} \sim 2.7$.

It is worth stressing that, as indicated above, the spectrum given by Eq.~(\ref{eq:effectiveSpectrum}) is formally that of a single source power-law distribution. Therefore, the propagated spectrum that would be obtained from such a model, after taking into account the interaction of the UHECRs with the intergalactic background fields, would fit the observed spectrum equally well, provided that the effective spectral index (Eq.~\ref{eq:xEff}) is adjusted in the same way as in the standard (single $E_{\max}$) model.

\begin{figure}
\begin{center}
  \includegraphics[angle=90,width=1.1\columnwidth]{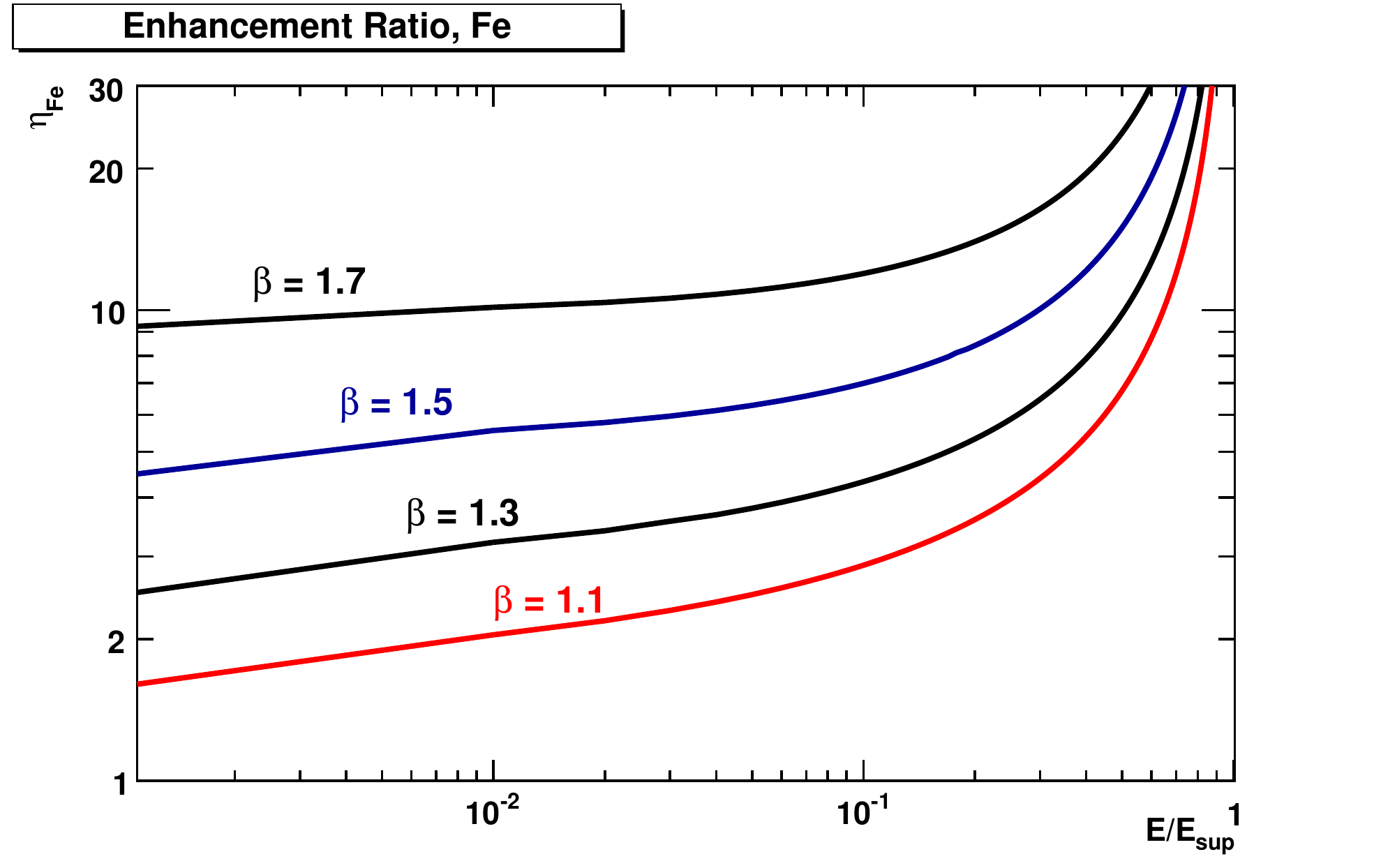}
\caption{The enhancement ratio of Fe nuclei, $\eta_{\mathrm{Fe}}$, Eq.~(\ref{eq:enhancementratio}), as a function of $E/E_{\sup}$, for several values of $\beta$, the slope of the $E_{\max}$ distribution.}
\label{fig:ratioplot}
\end{center}
\end{figure}

\section{The Abundance Enhancement Ratio at the Highest Energies}

At the very highest energies, when E becomes closer to $E_{\sup}$, the last factor in Eq.~(\ref{eq:effectiveInjection}) is no longer equal to $\sim 1$. The effective spectrum is then no longer a power-law, and the effective composition becomes dependent on energy.
We can define the enhancement ratio of the abundance of the various nuclei relative to protons, $\eta_{i}$, as
\begin{equation}
\label{eq:enhancementratio}
 \eta_{i}(E,\beta)\equiv \dfrac{q_{i}(E)/q_{\mathrm{p}}(E)}{\alpha_{i}/\alpha_{\mathrm{p}}} = \frac{Z_{i}^{\beta-1} - (E/E_{\sup})^{\beta-1}}{1 - (E/E_{\sup})^{\beta-1}}\,,
\end{equation}
where $q_{i}(E)$ was taken from Eq.~(\ref{eq:effectiveInjection}).

\begin{figure}
\begin{center}
\includegraphics[angle=90,width=1.1\columnwidth]{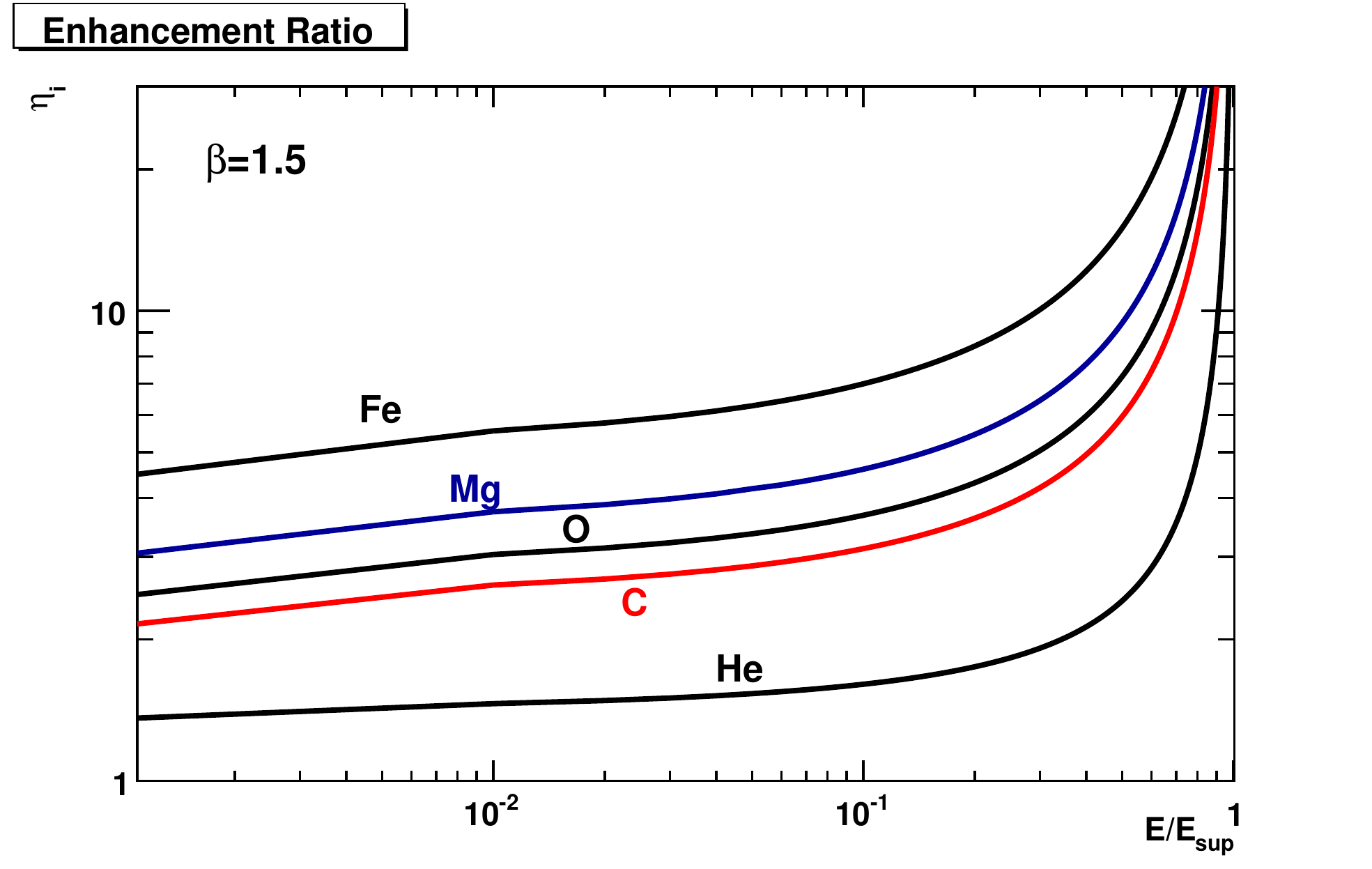}
\caption{The enhancement ratio, $\eta_{i}$, for various nuclei, as indicated, as a function of $E/E_{\sup}$.}
\label{fig:ratioplotNuclei}
\end{center}
\end{figure}

As an illustration, the behavior of the Fe enhancement ratio, $\eta_{\text{Fe}}$, as a function of $E/E_{\sup}$ is plotted in Fig.~\ref{fig:ratioplot} for several values of $\beta$. The low energy limit is given by $\eta_{i} = Z_{i}^{\beta-1}$ (see Eq.~(\ref{eq:alphaEff})). The enhancement ratio increases with energy. Likewise, the enhancement ratio is larger for heavier nuclei, as shown in Fig.~\ref{fig:ratioplotNuclei} for $\beta = 1.5$.

At the highest energies, close to $E_{\sup}$, the values shown on the plot should no longer be taken seriously, as in practice the granularity of the sources would have to be taken into account. The features of the cosmic ray energy spectrum and composition in this energy range are determined by the properties of the few highest energy sources, which are subject to ``cosmic variance''. It is obvious, however, that at energies above the proton maximum energy, $E_{\sup}$, the UHECRs should be completely devoid of protons, and thus the propagated UHECRs will be dominated by Fe and sub-Fe nuclei~\cite{Allard+08}.

\section{Conclusion}

We have studied the effect of a distribution of UHECR sources as a function of the maximum energy to which they can accelerate particles, both on the effective source spectrum and the effective source composition. Although the results shown here correspond to a specific set of assumptions, they show that the UHECR composition can indeed be significantly affected by such a distribution. These effects should thus be kept in mind when studying the phenomenology of UHECRs within the simplified framework of identical source models.

In particular, we showed that a heavier composition can be obtained very naturally under the astrophysically sensible assumption that the number of sources able to accelerate particles up to a given energy decreases sufficiently rapidly with that energy ($\beta > 1$). The abundance enhancement effect depends on the charge of the nuclei, and is less pronounced for lighter nuclei. This tends to enhance the contribution of Fe nuclei with respect to intermediate mass nuclei, which are also more affected by propagation effects due to their shorter horizon scales~\cite{Allard+08,Harari+06}.

Such a heavier composition is needed to fit the most recent UHECR data, within a model where a low value of $E_{\sup}$ is invoked to explain the apparent transition towards a heavy-nuclei-dominated component above a few $10^{19}$~eV. These models naturally account for a relatively sharp cut-off of the proton-dominated UHECRs, in the energy range where the sources cease to accelerate protons. However, if the heavier components are not sufficiently abundant in this energy range, this proton cut-off would show in the energy spectrum as a visible feature. Therefore, the consistency of such models requires a higher abundance of heavy nuclei than the typical expectations for the actual source composition.

As we have shown here, an enhancement ratio of the order of 3 for Fe nuclei appears natural if the $E_{\max}$ distribution function has a power law index $\sim 1.3$. This value of $\beta$ is needed to go from an intrinsic source spectrum of $E^{-2.0}$ to an effective source spectrum of $E^{-2.3}$, as is typically needed to fit the observed spectrum in the case of a mixed composition scenario~\cite{Allard+05,Allard+07b}. Larger enhancement ratios are possible if the $E_{\max}$ distribution function is steeper, which then implies that the individual source spectra are harder. A source spectrum of $E^{-1.8}$, for instance, would need a value of $\beta = 1.5$ to mimic a single-type source distribution with a spectrum of $E^{-2.3}$, resulting in an enhancement ratio of $\sim 5$ for Fe nuclei.

It should also be noted that, in the case of a low value of $E_{\sup}$, UHECR experiments are detecting cosmic rays just below and above $E_{\sup}$. The composition effects would then be expected to be \emph{larger} than discussed above (as apparent in Figs.~\ref{fig:ratioplot} and~\ref{fig:ratioplotNuclei}). In this transition range, across $E_{\sup}$, the analytical treatment used here is no longer relevant, and the local distribution of sources should be taken into account. For the same reason, a simple description of the overall spectrum in terms of power laws may not be possible for the highest energy cosmic rays. Even in the framework of the above continuous model, Eq.~(\ref{eq:effectiveInjection}) is no longer a power-law in this energy range, and one cannot define an effective power-law index as in Eq.~(\ref{eq:xEff}).

%The abundance enhancement effect discussed here depends on the charge of the nuclei, and is less pronounced for lighter nuclei. This tends to enhance the contribution of Fe nuclei with respect to intermediate mass nuclei, which are also more affected by propagation effects due to their shorter horizon scales (ref.).
We did not attempt a complete fit of the data here, as it will require additional assumptions about the individual source composition(s) and possibly some fine tuning of the parameters of the most nearby sources. However, in future works, we shall explore the cosmic variance around $E_{\sup}$ and investigate how the simple ideas studied here analytically can be applied to direct Monte-Carlo simulations with random source distributions, in order to create a global fit of present UHECR data, including both spectrum and composition.

%Although the $E_{\max}$ distribution function introduces an extra free parameter in the general phenomenology of UHECRs, it can be related to the astrophysical parameters of given source populations.
%Specifically, within a given UHECR source model, the value of $\beta$ can in principle be derived from physical considerations related to the acceleration model. As such, it enriches the ability to understand the UHECR phenomenon in an astrophysical context. In other words, this extra parameter allows for a relation between the parameters of UHECR phenomenology (derived from fits of the data within single identical source models) to more meaningful astrophysical parameters related to the actual acceleration mechanism and acceleration environment.



%% References


%\bibliographystyle{elsarticle-num}
%\bibliographystyle{unsrt}
%\bibliography{UHECRliterature.bib}



\end{document}